\documentclass{iauc}
\usepackage{graphicx}
\pubyear{2005}
\volume{199}
\pagerange{1--}
\jname{Probing Galaxies through Quasar Absorption Lines}
\editors{P. R. Williams, C. Shu, and B. M\'{e}nard, eds.}
\def\ltsima{$\; \buildrel < \over \sim \;$}
\def\simlt{\lower.5ex\hbox{\ltsima}}
\def\gtsima{$\; \buildrel > \over \sim \;$}
\def\simgt{\lower.5ex\hbox{\gtsima}}

\title[Absorption Lines and Feedback] %% give here short title %% 
{Galaxies, Intergalactic Absorption Lines, and Feedback at High Redshift}

\author[Kurt L. Adelberger]   %% give here short author list %%
{Kurt L. Adelberger$^1$}

\affiliation{$^1$Carnegie Observatories, 813 Santa Barbara St., 
Pasadena, CA 91101, USA \break 
email: kurt@ociw.edu}

\begin{document}

\maketitle

\begin{abstract}
The galaxy-IGM part of the Lyman-break survey currently consists of
measured redshifts for more than 1000 galaxies with redshift $1.5\simlt z\simlt 3.5$
along the sightlines to 25 background QSOs.  One of the goals of the
survey was to measure the influence on the intergalactic medium of 
energetic feedback from star and black-hole formation.   This talk 
begins with a description of
the observed correlations between galaxies and intergalactic absorption lines
and ends with a discussion of
whether any of the observations provide clear evidence for
Mpc-scale superwinds.  Although our own observations remain fairly
ambiguous, other observations strongly disfavor a very high redshift
($z\sim 10$) for the creation of intergalactic metals.
\keywords{galaxies: high-redshift, intergalactic medium, quasars: absorption lines}
%% add here a maximum of 10 keywords, to be taken form the file <Keywords.txt>.

\end{abstract}

\firstsection % if your document starts with a section,
\section{Introduction}
\begin{figure}
 \includegraphics{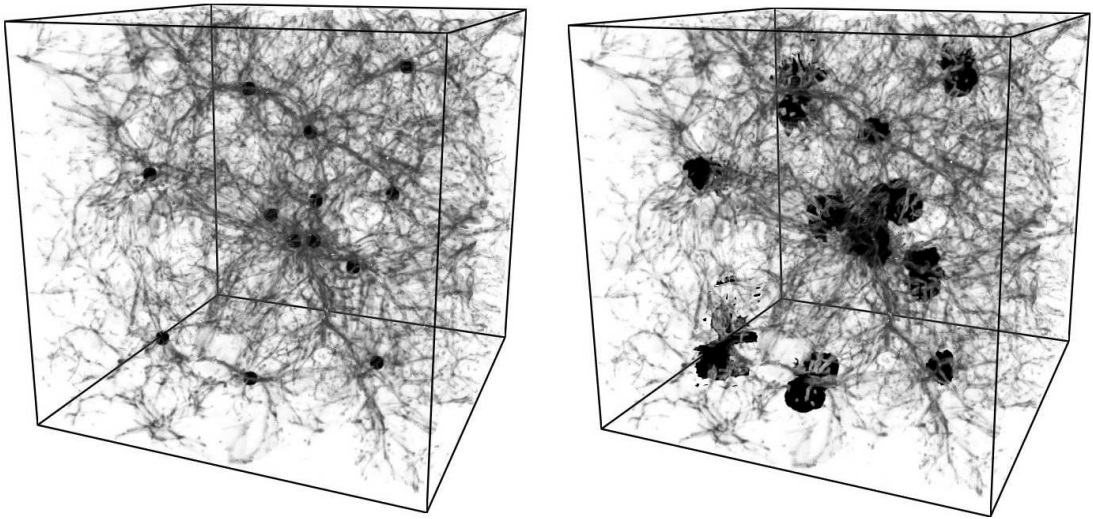}
  \caption{Cartoon illustration of the question that motivated us.  The filaments in the boxes
    show the distribution of gas in a numerical simulation of Cen \& Ostriker.
    Are galaxies (black spheres in left panel) surrounded by expanding superwinds,
    as illustrated by the large black regions in the right panel?
\label{fig:chen_intro}}
\end{figure}

Figure~\ref{fig:chen_intro} illustrates the subject of this talk in a simple
way.  We would like to determine which cartoon resembles the universe at high redshift
more closely.
The left panel shows data from Cen \& Ostriker's
numerical simulation of the evolving IGM.  Gas settles into a filamentary pattern, streams
into galaxies, and is peacefully converted into stars.  The right panel is similar,
except the energy released by each galaxy's supernovae and central black hole
creates an enormous blastwave that sends metal-enriched gas back into its surroundings.
Distinguishing between these possibilities is important, because standard models
of galaxy formation cannot reproduce many basic observations\footnote{Disk galaxies' sizes,
the shape of the luminosity function, the fraction of baryons that have been crushed 
into stars, the X-ray cluster temperature-luminosity relationship, the existence
of red galaxies at high redshift, etc.} unless something heats and disrupts
the gas near galaxies (although see Ari Maller's contribution for an alternate solution).
Establishing or ruling out the existence of ``superwinds'' would have far-reaching
implications.

One approach, taken by Rauch (this volume) and others, is to look for statistical differences
between the observed 
intergalactic absorption in QSO spectra and the predictions of a windless scenario.
A simple calculation shows the difficulty of this approach.  Suppose every Lyman-break galaxy (LBG)
at redshift $z\sim 3$ were surrounded by a spherical wind with a time-averaged radius
of $1h^{-1}$ comoving Mpc.  This is the maximum radius allowed by even the most optimistic
assumptions about the available energy (\cite[Adelberger 2002]{A02}).  Then the volume filling-fraction
of winds would be only 1.7\% for the observed LBG comoving number density of $4\times 10^{-3} h^3$
Mpc$^{-3}$.  Galaxies too faint to be selected as LBGs would presumably increase
the filling fraction somewhat, but still the impact of the tiny disturbed regions on the overall statistics of
the IGM would be small and difficult to discern.  This is especially true since 
absorption lines in the disturbed and undisturbed
regions trace the same large-scale structure and are both 
produced by gas that has settled to similar temperatures
of a few $\times 10^4$ K near the bottleneck in the cooling curve; their powerspectra
will therefore have nearly indistinguishable shapes in the large $k$ and small $k$ limits.

We adopted an alternate approach.
Since a weak signal is easier to detect if one knows where to look, we decided
to measure the redshifts of galaxies near QSO sightlines
and focus our attention on the $\sim 2$\% of the IGM that lay within
1 Mpc of an LBG.  Our sample currently consists of spectroscopic redshifts for
more than 1000 galaxies with $1.8\simlt z\simlt 3.5$ that lie in fields
containing 25 QSOs.  Over 500 CIV systems were detectable in our QSO spectra.

A detailed analysis of the survey can be found in \cite[Adelberger \etal\ (2005c)]{A05c}.  
Here is a brief summary of their main empirical results.  The interpretation
will be discussed afterwards.

$\bullet$ The gas that lies within 40kpc of LBGs produces extremely strong absorption
lines ($N_{\rm CIV}\gg 10^{14}$ cm$^{-2}$) in the spectra of background galaxies and QSOs.
The large equivalent widths of the CIV absorption in low-resolution spectra imply that the
absorbing material has range of velocities of at least $\Delta v=260$ km s$^{-1}$.
The absorption produced by this gas is similar to the interstellar absorption seen
in LBGs' spectra, suggesting the LBGs' outflowing ``interstellar'' gas may actually
lie at radii approaching 40kpc.

$\bullet$ For roughly half of the LBGs,
CIV absorption lines with equivalent width $N\sim 10^{14}$ cm$^{-2}$
are observed out to impact parameters of $\sim 80$ kpc.
This implies that roughly one-third of all ``intergalactic'' absorption lines
with $N\simgt 10^{14}$ cm$^{-2}$ are produced by gas that lies within 
$\sim 80$ kpc of an LBG.  Galaxies too faint to satisfy our selection criteria
could easily account for the remainder.  In some cases the absorbing gas
has substructure on half-kpc scales.

\begin{figure}
\includegraphics{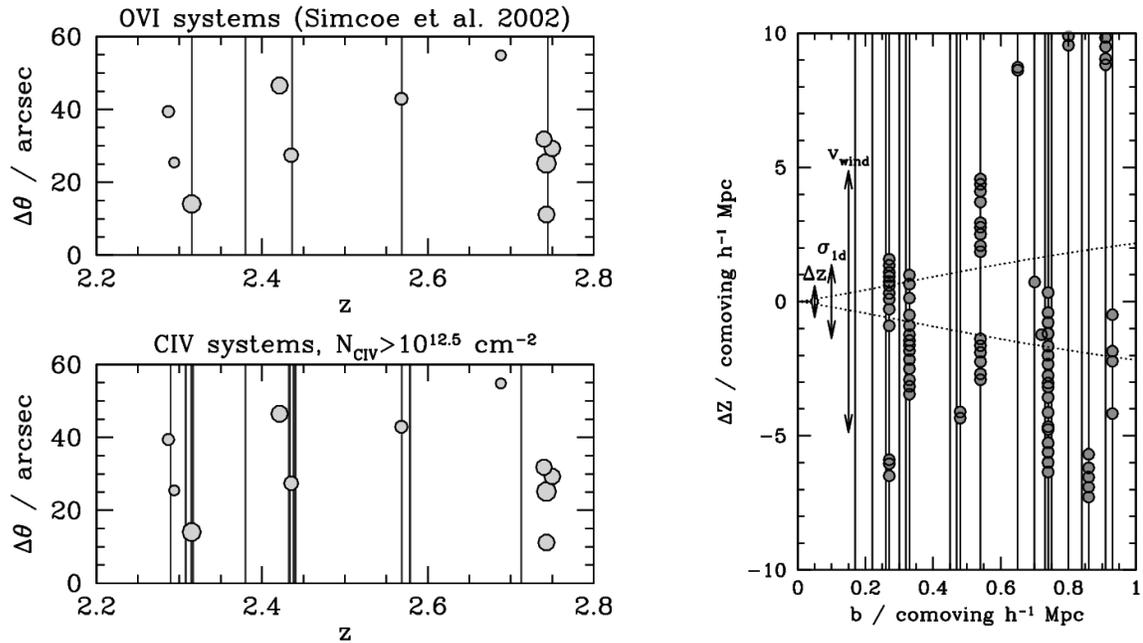}
  \caption{
Left panel:  
Redshifts of galaxies near the sightline
to QSO HS1700+6416 compared to the redshifts of
CIV and OVI absorption systems in the QSO's spectrum.
Circles mark galaxy redshifts and impact parameters.
The area of each circle is proportional to the galaxy's
apparent luminosity in the ${\cal R}$ band.
Vertical lines mark the redshifts of OVI and CIV absorption.
Thicker lines in the CIV panel mark systems with multiple components.
OVI absorption redshifts are taken from \cite[Simcoe, Sargent, \& Rauch (2002)]{S02}.
The absorption line catalogs are reasonably complete down to
their limiting equivalent widths, but,
owing to imperfect color-selection criteria and limited time for
follow-up spectroscopy, it is unlikely that the galaxy catalogs
contain more than half of the galaxies
brighter than ${\cal R}=25.5$ at these redshifts.
Right panel: 
The spatial separations of our sample's
galaxy-CIV pairs in the angular ($b$)
and redshift ($\Delta Z$) directions.  Vertical lines show the
projected distance between each galaxy in the NIRSPEC sample
and its nearest QSO.  The QSO spectra allow us to detect
any CIV-absorbing gas along these lines.  Circles
show the actual locations of detected CIV absorption.
Owing to the power-law shape of the galaxy-CIV correlation
function, half of the
circles (on average) would lie inside the dotted triangular envelope
if there were no peculiar velocities.
In fact 65 of 81 lie outside.
The uncertainty in each galaxy-CIV pair's redshift separation (i.e.,
in each circle's vertical position) is too small
to account for this ($\sim 60$ km s$^{-1}$ or $0.6h^{-1}$ comoving Mpc).
Peculiar velocities must therefore be substantial.
If the CIV systems were orbiting in the
galaxy's halo, their observed redshift separations would be displaced
relative to their true separations by a random amount comparable
to their galaxy's velocity dispersion $\sigma_{1d}$
($140$ km s$^{-1}$ or $1.4h^{-1}$ comoving Mpc). 
If the CIV were infalling, one would expect the CIV systems' observed
(i.e., redshift space) positions to be artificially compressed towards
the $x$ axis by an amount comparable to $\sigma_{1d}$.
If it were flowing outwards, the observed positions would be displaced
away from the $x$ axis by an amount that depends on the wind speed.
($\sim 500$ km s$^{-1}$ or $4.9h^{-1}$ comoving Mpc).
\label{fig:q1700civxigc}}
\end{figure}

$\bullet$  The cross-correlation function of galaxies and CIV systems with
$N_{\rm CIV}\simgt 10^{12.5}$ cm$^{-2}$
appears to be the same as the correlation function of galaxies.
This implies that CIV systems and galaxies reside in similar parts of
the universe and is consistent with the idea that they are largely
the same objects.  We also find a strong association of OVI systems with 
galaxies (upper left panel, Figure~\ref{fig:q1700civxigc}).
On small scales the 
redshift-space cross-correlation function is highly anisotropic 
(right panel, Figure~\ref{fig:q1700civxigc}).
The metal-enriched gas must therefore have large velocities
relative to nearby galaxies.  The required velocities appear to
exceed the galaxies' velocity dispersions 
but are similar to the galaxies' observed outflow speeds.

$\bullet$ In contradiction to the earlier result of~\cite[Adelberger et al. (2003)]{A03}, 
we find that the gas within $1h^{-1}$ comoving Mpc of LBGs 
usually produces strong Ly-$\alpha$ absorption in the spectra
of background galaxies.  The absorption is weak (mean transmitted flux 
within $1h^{-1}$ Mpc of $\bar f_{1{\rm Mpc}}>0.7$)
in only about one case out of three. % (Figure~\ref{fig:compare_juna}).  
Even so, the weakness of the absorption
in these cases remains difficult to understand.  Since high-redshift galaxies reside in
dense parts of the universe, with large amounts of hydrogen and short
recombination times, one would expect them to be surrounded by
large amount of HI---unless something had disrupted the nearby material.
One might imagine that the HI has collapsed into clouds and that the sightlines
to the background QSOs occasionally miss every cloud near a galaxy, 
but the SPH simulations of Kollmeier et al. (in preparation) suggest that
the chance of this is very low.% (Figure~\ref{fig:compare_juna}).

%\begin{figure}
%\includegraphics[height=3in]{compare_juna.eps}
%  \caption{
%Observed mean transmissivity of the IGM within $1h^{-1}$ comoving Mpc of
%LBGs compared to the predictions of a windless SPH simulation.  The
%observed mean transmissivity for each galaxy within $1h^{-1}$ comoving Mpc
%of the QSO sightline is shown in lighter bars; the darker histogram in
%the background shows the expected distribution for random skewers in
%the windless SPH simulation of Kollmeier et al. (2005; in preparation).
%\label{fig:compare_juna}}
%\end{figure}

\section{Discussion}
Galaxies would presumably be associated with QSO absorption lines even
if there were no winds.  I thought it might be useful to spend some
time discussing
how strongly (or weakly) the different observations
support the superwind scenario.  Table~\ref{tab:reportcard}
presents my views in the form of a report card.  
Each observation is graded on two criteria:  (a) how certain we
are, statistically, that the observation is correct,
and (b) how cleanly the observation would support the wind
scenario if it were correct.  I am adopting the Harvard grading
system, in which `A' is the highest grade you can get and everyone
gets it unless they are completely abject.

Although it was not the subject of my talk, I'll begin with the
blueshifted absorption and redshifted Lyman-$\alpha$ emission
that has been detected in hundreds of high-redshift galaxies
(e.g., \cite[Pettini et al. 2001]{P01}; \cite[Shapley et al. 2003]{Shap03};
Erb et al. 2005, in preparation).
This is expected in a wide range of models in which the stars
are surrounded by outflowing gas (e.g., \cite[Tenorio-Tagle et al. 1999]{TT99},
\cite[Zheng \& Miralda-Escude 2002]{ZH02}) and (as far as I know)
is not expected in any other class of realistic models.
Since it also cannot be a statistical fluctuation, I give it an A in both
columns.  There is a significant caveat, however:  it tells us only
that outflowing gas lies somewhere outside the stellar radius,
not that the outflow will spill out of the galaxy's
halo and qualify as a superwind.

The strength of CIV absorption at $b=40$ kpc may suggest
that the outflows have normally advanced to this radius,
but even so this is only about half way to the virial radius.
The minimum allowed velocity range of the gas at $b=40$ kpc
($\sim 300$ km s$^{-1}$) is not large enough to guarantee escape.
In any case the similarity of the CIV absorption strength in LBGs'
spectra and at impact parameter $b=40$ kpc could just be a coincidence; it does not
rule out the idea that the gas at 40kpc is falling into or
orbiting within the galaxy's potential.
Anisotropies in the galaxy/CIV-system correlation function disfavor
infall and seem to favor rapid outflows for the origin of the CIV,
but the small number of distinct galaxy-CIV pairs leaves the situation unclear.
All of these results are based on only a handful of galaxies
and their statistical significance merits only a C+.
In larger sample they could prove to be powerful diagnostics, however,
especially the correlation-function anisotropies and the changes
in CIV mean velocity and velocity width as a function of impact parameter.
That is the justification for the B+ in the interpretation column.

Measuring the spatial correlation of galaxies and metals on large (Mpc) scales
provides a powerful way to distinguish between different 
scenarios for intergalactic metal enrichment.  If the metals
were produced in LBGs, they should have a spatial bias similar to the
galaxies' bias of $b\sim 2.5$ at $z=3$ \cite[(Adelberger et al. 2005)]{A05a}.
They would have $b=1$ if they were produced at any redshift by 
the numerous dwarf galaxies that are $1\sigma$
fluctuations, and $b\sim 1.9$ at $z=3$ if (as envisioned by \cite[Madau, Ferrara, \& Rees 2001]{MFR01})  
they were produced at $z=9$ by galaxies that were $2\sigma$ fluctuations.
These estimates of the bias exploit \cite[Mo \& White's (1996)]{MW96}
high-redshift (i.e., $\Omega\sim 1$) approximation 
$b = 1 + (\nu^2-1)(1+z)[1.69(1+z_i)]^{-1}$ for the bias
at redshift $z$ of objects that were $\nu$-sigma fluctuations
at the earlier redshift $z_i$.
The galaxy-metal cross-correlation length would therefore be equal to the galaxy-galaxy
correlation length if the metals were produced by LBGs and smaller
for the other two cases.  Unfortunately we have measured the galaxy-CIV correlation length,
not the galaxy-metal correlation length.  Understanding the relationship
between the two will likely
require sophisticated numerical simulations that may not be available
for many years.  Since our galaxy-CIV clustering measurements will
not sway the skeptical until that time, I have assigned them a C
for their interpretive value.  They merit no more than a B for significance
since our $2\sigma$ confidence intervals would allow galaxy-CIV
clustering to be considerably weaker than the galaxy-galaxy clustering.

\begin{table}\def~{\hphantom{0}}
  \begin{center}
  \caption{Report Card}
  \label{tab:reportcard}
  \begin{tabular}{lll}\hline
      Observation  & Statistical Significance &  Interpretation \\\hline%\\%[3pt]
       Blueshifted ISM, redshifted Ly$\alpha$  & A  & A \\
       Metals at $r\simlt 100$ kpc             & C+ & B+\\
       Metals at large separations             & B  & C \\
       Lack of HI near some galaxies           & C+ & C+\\\hline
  \end{tabular}
  \end{center}
\end{table}

Although there are now 9 galaxies in our sample with precise (near-IR nebular line)
redshifts and little HI within $1h^{-1}$ comoving Mpc, their statistical
significance is worth only a C+ in my opinion.  There are two reasons.
First, the measurement is difficult.  To determine
whether a galaxy lies within $\simlt 1h^{-1}$ Mpc of a narrow absorption line
in a QSO spectrum, one needs
a good understanding of many possible sources of random and systematic error
in the estimated redshifts.
In about half the cases there is no absorption line anywhere near the galaxy redshift,
but in the remainder the galaxies could conceivably be associated with nearby 
absorption lines
if redshift errors were somewhat larger than we estimate.
Reducing the number of galaxies with little nearby HI by 50\% would
begin to make our observations consistent with the predictions of windless SPH simulations.
Second, 5 of the 9 galaxies lie within a single field, Q1623, which contains one of
the largest known concentrations of QSOs with $2\simlt z\simlt 3$.  
This is a consequence of the dense spectroscopic sampling we obtained in the field,
but it raises the possibility that our result might have been different
had we surveyed a more representative part of the universe.
The spatial clustering strength of the high-redshift galaxies
in this field is certainly not typical, for example (\cite[Adelberger et al. 2005b]{A05b}).

Interpreting the result is difficult because of the complexity of the any interaction
between winds and galaxies' inhomogeneous surroundings.
This cannot be modeled analytically, and
existing numerical simulations are unable to resolve either
the shock fronts or the instabilities that result when the hot wind
flows past cooler intergalactic material.  As a result it is difficult
to know whether winds would destroy the HI near galaxies.
Other effects that simulations do not resolve (e.g., cooling
instabilities) might reduce the covering fraction of HI near galaxies.
Even if there were low-density regions near galaxies in SPH simulations,
they might not be recognized since the density is generally estimated
by smoothing over the few dozen nearest particles, particles which
(by definition) are likely to lie outside of any local underdensity.
It therefore remains uncertain whether the lack of HI near some galaxies is
necessary or sufficient to establish the existence of superwinds.
The lack of any correlation between galaxy properties and
the strength of nearby intergalactic HI absorption (\cite[Adelberger et al. 2005c]{A05c})
makes me suspect
that the weak observed HI absorption is 
probably unrelated to winds.
I have given this observation a ``C+'' for interpretative value,
but that is generous.

\section{Concluding Remarks}
It is easy to measure the relative spatial positions of galaxies and
the gas that produces QSO absorption lines, but difficult to find
conclusive evidence for superwinds around galaxies.  This is partly
because our sample is still small, but mostly, I think,
because theorists have been unable to work out the observational
consequences of superwinds.
It is entirely possible, for example, that our measured galaxy-CIV correlation
function rules out every explanation except superwinds,
but we would not know since the necessary simulations do not exist.

\begin{figure}
\includegraphics[height=3in]{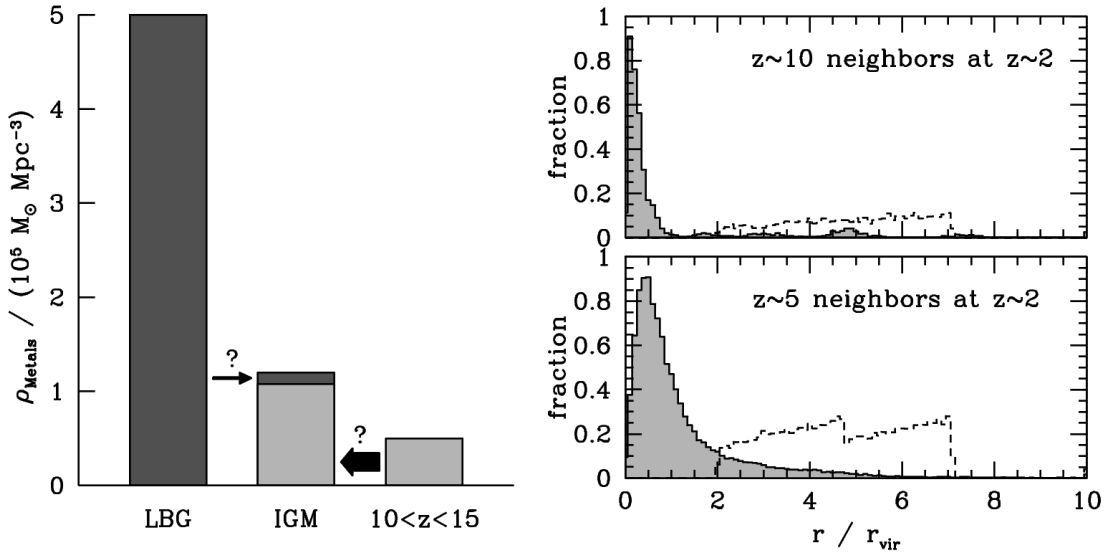}
  \caption{
\textit{Left panel:}
The amount of metals produced in massive galaxies at $z\sim 2$ (LBG) compared to
the amount of metals in the IGM at $z\sim 2$ and to the amount of
metals produced at $10<z<15$.  The amount of metals produced by massive galaxies
at $z\sim 2$ was calculated by assuming that each $100 M_\odot$ of star formation
produced $1 M_\odot$ of metals and scaling from the $z\sim 2$ stellar-mass
density measured by \cite[Dickinson et al. (2003)]{D03}.  An upper limit
to the amount of metals produced at $10<z<15$ was derived by assuming a constant
comoving star-formation density at all redshifts $2<z<15$.  This implies
that $\sim 10$\% of the stars that exist at $z=2$ were formed
at $10<z<15$.  It is an upper limit because the actual star-formation density
at $z\gg 2$ appears to be significantly lower than the star-formation density
at $z\sim 2$.  The amount of metals in the IGM was calculated by multiplying
the critical density by the estimated intergalactic metal density at $z\sim 2$, 
$\Omega_{\rm met}\sim 4.4\Omega_C$ with $\Omega_C\sim 2\times 10^{-7}$
(\cite[Schaye et al. 2003]{Sch03}).
If 90\% of the intergalactic metals at $z\sim 2$ were produced at $10<z<15$,
the fraction of metals that escape into the IGM would have to be
$\sim 100$ times higher at $10<z<15$ than at $z\sim$2--3.
\textit{Right panels:}
Distance to the nearest massive ($M\simgt 10^{11}M_\odot$)
halo at $z\sim 2$ for all GIF-LCDM simulation
particles whose distance $r$ to the nearest halo satisfied
$2r_{\rm vir} < r < 1h^{-1} {\rm comoving Mpc}$ at $z=10$ (top) or $z=5$
(bottom).  Dashed lines show the particles' original 
(higher redshift) distribution
of $r/r_{\rm vir}$; the solid shaded histograms show the
distribution at $z\sim 2$.  These particles initially lie
at larger radii than those expected for the metals ejected 
by very high redshift winds, yet they mostly end up inside halos
by $z\sim 2$.  This implies that the metals ejected at $z\sim 5$ and $z\sim 10$
will generally also lie in massive halos at $z\sim 2$, not in the IGM.
\label{fig:metals}}
\end{figure}

Although I have emphasized the weaknesses in the superwind interpretation,
I should point out that in my opinion the weaknesses in the alternate scenario
are far worse.  
The left panel of Figure~\ref{fig:metals} illustrates the major problem faced by those
who believe that intergalactic metals were created in early
episodes of star formation at $10\simlt z\simlt 15$:  
they must assume that the metals produced
at $z\sim 10$ escape from their galaxies with ease
while almost none of the metals produced at $z\sim 3$
are able to do so.  In order for 90\% of the 
intergalactic metals observed at $z\sim 3$ to have
been produced at $z\sim 10$, for example, 
supernovae ejecta would have to be at least 100 times
more likely to escape their galaxy at $z=10$
than at $z=3$.  (See the caption to Figure~\ref{fig:metals}
for the derivation of this number.)  
It is widely believed that high-redshift supernovae ejecta escape 
more easily because galaxy
masses are lower at high redshift, and that they pollute the IGM more readily
because they need to reach only a comparatively small physical radius
to enrich a large comoving volume.  \cite[Scannapieco (2005)]{Sc05}
argues that these two effects cause the comoving filling fraction of the ejecta $Q$
(or, more formally, the porosity) to depend on mass and redshift
as $Q\propto M^{-2/5}(1+z)^{6/5}$.  (Similar equations have
been derived by Madau et al. 2001 and dozens of other papers.)
This equation predicts that
metals from $10^9M_\odot$ galaxies at $z=10$ should have a filling
fraction $\sim 20$ times larger than those from the $10^{11}M_\odot$
galaxies observed at $z\sim 3$.  It is not clear to me whether
this is enough to explain the left panel of Figure~\ref{fig:metals}, but the question
is moot:  in my view, \cite[Scannapieco's (2005)]{Sc05} formula vastly overstates
the actual advantage of very high-redshift galaxies.

There are two reasons.  First, it assumes that small galaxies at high redshift
convert their baryons into stars exactly as efficiently as more massive galaxies
at $z\sim 3$.  This assumption is probably wrong by an order of magnitude.
Star-formation efficiencies in massive galaxies at $z\sim 3$ appear to be far higher,
as is easily seen by comparing the ratio of stellar to baryonic mass
in LBGs ($M_\ast/M_b \sim 0.1$; \cite[Adelberger et al. 2004]{A04})
to the global average at $z=3$, $\Omega_\ast / \Omega_b\simeq 0.01$ 
(\cite[Dickinson et al. 2003]{D03}; \cite[Rudnick et al. 2003]{R03}).
Since every baryon at $z=3$ was part of a low-mass halo at higher redshift,
at least in the Press-Schechter picture, and only $\sim 1$\% of them had been
turned into stars by $z=3$, the mean star-formation efficiency in low mass halos at high
redshift is evidently $\sim 10$ times lower than the mean efficiency in LBGs.  
A more realistic comparison would omit the halos with masses $M\simlt 10^5M_\odot$
that are too small to host star formation and would note that $\Omega_\ast/\Omega_b$
is an order of magnitude lower at $z\sim 10$ than $z\sim 2$ for
realistic cosmic star-formation histories.  This leads to a similar conclusion.
The high efficiency of star-formation in massive galaxies at $z\sim 2$ removes
the low-mass halos' supposed advantage in enriching the IGM.
Second, only a fraction of the metals released into the IGM at $z\sim 10$ will
remain in the IGM at $z\sim$2--3. 
This is a simple consequence of hierarchical
structure formation:  the material that lies near galaxies at higher redshift
(e.g., any metal-enriched ejecta)
will lie ${\it inside}$ galaxies at lower redshift.
Because they neglect this, standard treatments
seriously overestimate the intergalactic volume that metals produced at
$z\sim 10$ will occupy at $z\sim 2$. 
To illustrate the point, I calculated
the distance to the nearest halo of mass $M\simgt 10^{11} M_\odot$
at $z=2.12$ for every particle
in the GIF-LCDM simulation (\cite[Kauffmann et al. 1999]{K99})
that lay within $1h^{-1}$ comoving Mpc of a galaxy (i.e., halo)
at $z=10$ or $z=5$.  The result, shown in the right hand panels of Figure~\ref{fig:metals},
is stark:  the overwhelming majority of these particles end up inside
the virial radius of a galaxy at $z\sim 2$.  Since the stalled
ejecta of galaxies at $z=10$ or $z=5$ will be swept along by the
movements of the material that surrounds the galaxies, it should
largely end up inside galaxies at $z\sim 2$ as well.
This exercise may be slightly misleading,
since the GIF-LCDM simulation only resolves halos of mass $M\simgt 10^{11}M_\odot$,
not the smaller halos believed to be most responsible for polluting the
IGM at $z\sim 10$.  However, a simple calculation shows that a significant fraction
of the metals from the lower-mass progenitors should also end up inside galaxies
at $z\sim 2$.  The estimated stalling radius for winds from small galaxies 
at $z\sim 10$ is $\sim 100$ comoving kpc (e.g., \cite[Madau et al. 2001]{M01}), which
is the Lagrangian radius for a halo of mass $1.6\times 10^8 M_\odot$.  If these galaxies'
typical descendants at $z\sim 2$ have masses significantly larger than this,
the metals will likely have been swept inside them.  According to the extended
Press-Schechter formalism,
$\sim 85$\% of the galaxies at $z=2$
that descended from $2\sigma$ fluctuations at $z=10$ will have masses that exceed
this threshold by an order of magnitude
(see, e.g., equation 2.16 of \cite[Lacey \& Cole (1993)]{LC93}).  A substantial fraction of the metals
produced at $10\simlt z\simlt 15$ should therefore be locked inside galaxies by $z\sim 2$.
Once there, some of the metals are likely to cool further, fall towards the center,
and disappear from view.
Since the total
metal production at $10\simlt z\simlt 15$ is at best comparable
to the observed metal content of the IGM at $z\sim$2 (left panel, Figure~\ref{fig:metals}),
this suggests that the intergalactic metallicity at $z\sim 2$ must receive a significant
contribution from some other source.
A corollary is that the metal content of the IGM would drain into galaxies and
{\it decrease} over time if it were not continually replenished.  The observed
constancy of the IGM metallicity (e.g., \cite[Schaye et al. 2003]{S03}) therefore also seems to require
metals to escape from galaxies at lower redshifts.

In addition, a host of indirect arguments point towards superwinds from relatively massive
galaxies at $z\sim 3$.
Clustering measurements show that galaxies that were forming star rapidly at $z\sim 3$
are no longer forming stars by $z\sim 1$ (\cite[Adelberger et al. 2004]{A04}).
What physical process is responsible for this, if not energetic feedback
that strips away the galaxies' gas and slows subsequent accretion?
If winds are only able to escape
from dwarf galaxies, how come so large a fraction of metals in the local universe are
found in the intracluster medium, near the largest galaxies?
I could carry on but my allotted time is up.
Although we have not yet found unequivocal evidence that intergalactic metals
at $z\sim$ 2--3 are often produced by galaxies at similar redshifts,
the burden of proof surely falls on the other camp.  Theirs
is the preposterous claim.

\begin{acknowledgments}

It's a pleasure to acknowledge several helpful
conversations with T. Abel, I. Labb\'e, P. Madau, 
M. Rauch, 
D. Weinberg, S. White, and my collaborators on the Lyman-break survey.
I am grateful to the organizers for the assistance
they provided before, during, and (especially) after
the meeting.  This work was supported by a fellowship
from the Carnegie Institute of Washington.
\end{acknowledgments}

\end{document}